\documentclass[twocolumn,aip,jcp,reprint,superscriptaddress,showpacs,floatfix]{revtex4-1}
\usepackage[latin9]{inputenc}
\usepackage{color}
\usepackage{longtable}
\usepackage{bm}
\usepackage{amsmath}
\usepackage{amssymb}
\usepackage{graphicx}
\usepackage{esint}

\makeatletter

\providecommand{\tabularnewline}{\\}

\makeatother

\begin{document}
\title{Improved estimation for energy dissipation in biochemical oscillations}
\author{Zhiyu Cao}
\author{Zhonghuai Hou}
\thanks{E-mail: hzhlj@ustc.edu.cn}
\affiliation{Department of Chemical Physics \& Hefei National Laboratory for Physical
Sciences at Microscales, iChEM, University of Science and Technology
of China, Hefei, Anhui 230026, China}
\date{\today}
\begin{abstract}
Biochemical oscillations, regulating the timing of life processes,
need to consume energy to achieve good performance on crucial functions,
such as high accuracy of phase period and high sensitivity to external
signals. However, it is a great challenge to precisely estimate the
energy dissipation in such systems. Here, based on the stochastic
normal form theory (SNFT), we calculate the Pearson correlation coefficient
between the oscillatory amplitude and phase, and a trade-off relation
between transport efficiency and phase sensitivity can then be derived,
which serves as a tighter form than the estimator resulting from the
conventional thermodynamic uncertainty relation (TUR). Our findings
demonstrate that a more precise energy dissipation estimation can
be obtained by enhancing the sensitivity of the biochemical oscillations.
Moreover, the internal noise and amplitude powers effects have also
been discovered.
\end{abstract}
\maketitle

\section{Introduction}

To achieve good performance of certain functions, living systems are
inherently nonequilibrium and dissipative. Recently, the relationship
between biochemical functions and nonequilibrium thermodynamics has
been an active area in statistical physics community \citep{bialek2005physical,hu2010physical,lan2012energy,lan2013cost,skoge2013chemical,lang2014thermodynamics,sartori2015free,cao2015free,fei2018design,mora2019physical,zhang2020energy,barato2016cost,lee2018thermodynamic,hasegawa2018thermodynamics,hasegawa2019uncertainty,marsland2019thermodynamic,del2020high,guan2020trade,del2020robust,cao2021designing}.
For instance, Lan $et$ $al.$ have revealed a powerful trade-off
relation between energy dissipation rate, adaption speed and the maximum
adaption accuracy underlying many sensory systems \citep{lan2012energy,lan2013cost}.
Lang $et$ $al.$ have investigated the fundamental thermodynamic
constraints on statistical inference and learning of biochemical signaling
networks \citep{lang2014thermodynamics}. Particularly, for biochemical
oscillations which are essential in regulating the timing of life
processes, such as the cell cycle, circadian clocks, and glycolysis,
both accuracy of the period and sensitivity to external cues can be
ensured by dissipative processes simultaneously \citep{ferrell2011modeling,buzsaki2004neuronal,nakajima2005reconstitution,novak2008design,goldbeter1997biochemical,martiel1987model}.
Therefore, it is important to measure the free energy dissipation
in biochemical oscillation systems that maintains the cyclic dynamics.
However, in actual experiments, how to infer the energy dissipation
is of great challenge \citep{gnesotto2018broken,seifert2019stochastic}.

Recent progress in this topic is the thermodynamic uncertainty relation
(TUR) \citep{barato2015thermodynamic,gingrich2016dissipation,pietzonka2016universal,pietzonka2017finite,dechant2018entropic,dechant2018current,agarwalla2018assessing,hasegawa2019fluctuation,horowitz2020thermodynamic},
quantifying the trade-off between energy dissipation $\Delta W$,
the average $\left\langle R\right\rangle $ and variance $\text{Var}\left(R\right)=\left\langle (R-\left\langle R\right\rangle )^{2}\right\rangle $
of a time-integrated current observable $R$ in nonequilibrium steady
states (here $\beta={\color{red}{\color{black}1/k_{B}T}}$, $T$ is
the temperature of the environment and $k_{B}$ is the Boltzmann constant):

\begin{equation}
\eta(R)=\frac{2\left\langle R\right\rangle ^{2}}{\text{\ensuremath{\beta}Var}(R)\Delta W}\le1,
\end{equation}
where $\eta(R)$ is the transport efficiency to properly quantify
the performance of living systems working with high accuracy, but
low energy dissipation \citep{dechant2018current}. Directly, TUR
yields that the magnitude of current fluctuation provides a lower
bound of energy dissipation as $\Delta W\ge\Delta W_{\text{TUR}}\equiv2k_{B}T\left\langle R\right\rangle ^{2}/\text{Var}(R)$
with $\eta\left(R\right)=\Delta W_{\text{TUR}}/\Delta W$. If $\eta(R)$
is close to $1$, the TUR acts as a powerful tool for energy dissipation
inference \citep{li2019quantifying,manikandan2020inferring,van2020entropy,otsubo2020estimating,skinner2021improved,skinner2021estimating,dechant2021continuous}.
For instance, recently Li $et$ $al.$ have showed that the fluctuations
in nonequilibrium currents can be utilized to infer the dissipation
rate for the bead-spring model \citep{li2019quantifying}. Otsubo
$et$ $al.$ have developed a framework for dissipation estimation
by using the TUR along with machine learning techniques \citep{otsubo2020estimating},
to list just a few.

However, since the TUR is an inequality, only a rough bound can be
provided in many cases. For instance, it has been revealed by Hwang
and Hyeon that the TUR is generally not tight for several types of
molecular motors \citep{hwang2018energetic}. Jack $et$ $al.$ have
found that the TUR only yields a weak bound for molecular-scale energy
conversion \citep{jack2020thermodynamic}. Also, in our recent work
\citep{cao2020design}, we have established the TUR for general biochemical
oscillations by calculating the transport efficiency $\eta(\theta)=2\left\langle \theta\right\rangle ^{2}/\text{\ensuremath{\beta}Var}(\theta)\Delta W<1$,
where the observable oscillatory phase $\theta(\tau)=\int_{0}^{\tau}\dot{\theta}(t)dt$
is the current observable. Both the analytical and numerical results
have shown that the TUR is far from tight for models of chemical oscillators,
providing typically lower estimation for energy dissipation than the
actual value. Therefore, how to obtain a more qualified estimation
than the conventional TUR for biochemical oscillation systems is still
an interesting question.

In the presented paper, we try to address this question by revealing
a trade-off relation between transport efficiency and phase sensitivity
\citep{hasegawa2014circadian,hasegawa2014optimal,fei2018design}.
The basic idea is to improve the conventional TUR by considering the
Pearson correlations between the chosen current and another state-dependent
observable, based on a strategy proposed by Dechant and Sasa very
recently \citep{dechant2021improving}. For practical purpose in biochemical
oscillation systems, we choose the time integral of the oscillatory
amplitude $r$ as the state-dependent observable, which reads $Q_{2}(\tau)=\int_{0}^{\tau}r^{2}(t)dt$.
By using the stochastic normal form theory (SNFT) we established before
\citep{hou2006internal,xiao2007effects,ma2008coherence,xiao2009stochastic},
explicit theoretical expressions of the Pearson correlations between
$Q_{2}$ and $\theta$ can be derived, which allows us to obtain the
efficiency-sensitivity trade-off relation as $\eta(\theta)\le1-2\alpha\kappa^{2}$
with $\kappa$ the phase sensitivity characterizing the ability for
biochemical circuits to respond to external signals and $\alpha>0$
the control parameter denoting the distance to the bifurcation point.
Remarkably, this trade-off relation provides a tighter dissipation
estimator for biochemical oscillations than the conventional TUR,
and the precision of this estimator can be further improved by enhancing
the sensitivity. Finally, we demonstrate our statements by detailed
numerical simulations in a circadian clock model.

\section{Improved estimation of the energy dissipation}

\subsection{Stochastic Normal Form Theory (SNFT)}

We consider a general biochemical system of size $V$ including $N$
well-stirred species and $M$ reactions as $(R_{1},\dots,R_{M})$.
Generally, the reaction $R_{\rho}$ can be written as:
\[
\mathbf{X}\to\mathbf{X}+{\bf v}_{\rho}
\]
where $\mathbf{X}=\left(X_{1},X_{2},\dots,X_{N}\right)$ with $X_{j}$
the number of species $j$, and $\mathbf{v}_{\rho}=\left(v_{\rho}^{1},v_{\rho}^{2},\dots,v_{\rho}^{N}\right)$
with $v_{\rho}^{j}$ the stoichiometric change of species $j$ in
$R_{\rho}$. In a mesoscopic system wherein intrinsic noise cannot
be neglected, with the assumption of existence of a ``macro-infinitesimal''
time scale \citep{gillespie2000chemical,xiao2007effects}, the system's
dynamics can be described by the chemical Langevin equations (CLEs)
as
\begin{equation}
\dot{x}_{j}=\sum_{\rho=1}^{M}v_{\rho}^{j}w_{\rho}(\bm{x})+\frac{1}{\sqrt{V}}\sum_{\rho=1}^{M}v_{\rho}^{j}\sqrt{w_{\rho}(\bm{x})}\xi_{\rho}(t),\:j=1,...,N.\label{eq:CLE}
\end{equation}
where $\bm{x}=(x_{1},\dots,x_{N})^{\text{T}}=\mathbf{X}/V$ denotes
the concentration vector, $w_{\rho}(\bm{x})$ is the reaction rate
of $R_{\rho}$ as a function of the concentrations $\bm{x}$, and
${\normalcolor \bm{\xi}(t)=(\xi_{1},\dots,\xi_{M})^{\text{T}}}$ is
a vector of independent Gaussian white noises with zero means and
correlations $\left\langle \xi_{\rho}(t)\xi_{\rho^{\prime}}(s)\right\rangle =\delta_{\rho\rho^{\prime}}\delta(t-s)$.

In the thermodynamic limit with $V\to\infty$, the noise term disappears
and the dynamics is described by the deterministic equation
\begin{equation}
\dot{x}_{j}=F_{j}\left(\bm{x}\right)\equiv\sum_{\rho=1}^{M}v_{\rho}^{i}w_{\rho}\left(\bm{x}\right)\label{eq:Deter_Eq}
\end{equation}
Generally, to the occurrence of biochemical oscillation, we assume
that the system undergoes a supercritical Hopf bifurcation (HB) with
the change of a certain control parameter $\mu$. Eq.(\ref{eq:Deter_Eq})
has a unique stable point $\bm{x}_{s}$ with $\bm{F}(\bm{x}_{s})\equiv0$,
which loses stability at the HB point $\mu=\mu_{c}$, in the way that
the Jacobian matrix $\bm{J}$ with components $J_{ij}=\left(\partial f_{i}/\partial x_{j}\right)|_{\bm{x}=\bm{x}_{s}}$
has a pair of conjugate eigenvalues $\lambda_{\pm}=\alpha(\mu)\pm i\omega$
with $\alpha(\mu_{c})=0$ (henceforth we uses $\alpha=\alpha(\mu)$
to represent the control parameter). In the so-called supercritical
region $\alpha>0$ ($\mu>\mu_{c}$), the deterministic system shows
a stable oscillation with frequency given by $\omega$ and amplitude
growing from zero. In the subcritical region with $\text{\ensuremath{\alpha<0}}$
($\mu<\mu_{c}$), no deterministic oscillation can be observed. In
the case where the system size is not large such that the internal
noise term in Eq.(\ref{eq:CLE}) can not be ignored, such as for intracellular
biochemical oscillation systems considered here, an interesting phenomenon
known as noise induced oscillations (NIOs) has been observed even
in the subcritical region where $\alpha<0$, demonstrating the constructive
role of internal noise in mesoscopic chemical oscillation systems
\citep{ko2010emergence}. In addition, an optimal system size exists
where the NIO shows best performance, knows as internal noise coherence
resonance (INCR) \citep{hou2003internal,zhou2002noise,hanggi2002stochastic}.

In our previous works \citep{hou2006internal,xiao2007effects,ma2008coherence,xiao2009stochastic},
we have developed a stochastic normal form theory (SNFT) to successfully
elucidate the mechanism underlying NIO and INCR. When the system locates
near the HB, the motion of the oscillatory mode is much slower than
the other $N-2$ stable modes due to time-scale separation. Hence,
the system's dynamics will be dominated by the oscillatory motion
on a 2D center manifold. According to SNFT, the stochastic dynamics
governing the evolution of the oscillation amplitude $r$ and and
phase angle $\theta$ can be described by (see Appendix A for details)
\begin{equation}
\dot{r}=\alpha r+C_{r}r^{3}+\frac{\varepsilon^{2}}{2Vr}+\frac{\varepsilon}{\sqrt{V}}\eta_{r}(t),\label{eq:dr}
\end{equation}

\begin{equation}
\dot{\theta}=\omega+C_{i}r^{2}+\frac{\varepsilon}{r\sqrt{V}}\eta_{\theta}(t)\label{eq:dth}
\end{equation}
wherein $C_{r}<0$ and $C_{i}>0$ are system-dependent constants determined
by the nonlinear terms of $\bm{F}\left(\bm{x}\right)$ at the stable
point, $\eta_{r}$ and $\eta_{\theta}$ are independent Gaussian white
noises with zero mean and unit variance, $\varepsilon$ denotes an
effective noise intensity determined by the details of $\bm{F}\left(\bm{x}\right)$.
According to Eqs.(\ref{eq:dr}) and (\ref{eq:dth}), the steady-state
(SS) distribution of $r$ reads
\begin{equation}
p_{ss}\left(r\right)=N_{r}\exp\left[-\frac{V}{4\varepsilon^{2}}\left(2\alpha r^{2}+C_{r}r^{4}\right)+\ln r\right]\label{eq:Phi}
\end{equation}
and $\theta$ is uniformly distributed with $\left[0,2\pi\right]$.
Therefore, the system exhibits a stochastic oscillation with most-probable
amplitude given by
\begin{equation}
r_{m}=\left(-\frac{\sqrt{\alpha^{2}-2C_{r}\varepsilon^{2}/V}+\alpha}{2C_{r}}\right)^{1/2}\label{eq:rm}
\end{equation}
satisfying $\partial p_{ss}\left(r\right)/\partial r|_{r_{m}}=0$.

Clearly, in the deterministic limit ($V\to\infty$), $r_{m}=\sqrt{-\alpha/C_{r}}$
corresponding to a stable limit cycle and frequency $\omega_{s}=\omega+C_{i}r_{m}^{2}=\omega+\alpha\left|C_{i}/C_{r}\right|$,
which only exists for $\alpha>0$ in the supercritical region. If
the system size is finite, however, the internal term $2C_{r}\varepsilon^{2}/V$
in the square-root will take effect and $r_{m}$ is not zero even
for $\alpha<0$ (subcritical region), corresponding to the occurrence
of NIO. In the case $\left|\alpha\right|\gg2C_{r}\varepsilon^{2}/V$,
one has for NIO $r_{m}\simeq\varepsilon/\sqrt{-2\alpha V}$ which
scales as $V^{-1/2}$, and the frequency is approximately $\text{\ensuremath{\omega_{s}}}\simeq\omega+C_{i}\varepsilon^{2}/\left(2\left|\alpha\right|V\right)$.
Therefore,
\begin{equation}
\omega_{s}=\begin{cases}
\omega+\alpha\left|C_{i}/C_{r}\right| & \left(\alpha>0\right)\\
\omega+C_{i}\varepsilon^{2}/\left(2\left|\alpha\right|V\right) & \left(\alpha<0\right)
\end{cases}\label{eq:Omega_S}
\end{equation}

\subsection{Transport Efficiency and Phase Sensitivity}

The purpose of the present work is to figure out a way to improve
the estimation of energy dissipation (or entropy production) related
to the stochastic oscillations. As mentioned in the introduction,
one usually uses the thermodynamic uncertain relation (TUR) as an
inference of the real energy dissipation via $\Delta W\ge\Delta W_{\text{TUR}}\equiv2k_{B}T\left\langle R\right\rangle ^{2}/\text{Var}(R)$
wherein $R$ is some well-defined current variable, and $Var\left(R\right)=\left\langle R^{2}\right\rangle -\left\langle R\right\rangle ^{2}$
denotes the variance of $R$. Correspondingly, the transport efficiency
for $R$ reads $\eta\left(R\right)=\Delta W_{\text{TUR}}/\Delta W=2k_{B}T\left\langle R\right\rangle ^{2}/\text{Var}(R)\le1$.
For the oscillatory dynamics considered here, it is convenient to
choose $R$ as the change of phase angle within a given time interval
$\left(0,\tau\right)$ , i.e., $R\left(\tau\right)\to\theta\left(\tau\right)=\int_{0}^{\tau}\dot{\theta}\left(t\right)dt$.
By simply rewriting and setting $k_{B}T=1$ from now on, the transport
efficiency can be expressed as $\eta(\theta)=2k_{B}T\left\langle \theta\right\rangle ^{2}/\text{\ensuremath{[}Var}(\theta)\Delta W]=v_{\theta}^{2}/D_{\theta}\dot{W}$
where $v_{\theta}=\lim_{t\to\infty}\left\langle \theta\right\rangle /t$
is the phase speed, $D_{\theta}=\lim_{t\to\infty}(\left\langle \theta^{2}\right\rangle -\left\langle \theta\right\rangle ^{2})/2t$
is the phase diffusion constant, and $\dot{W}=\lim_{t\to\infty}\Delta W/t$
is the dissipation rate.

By using the SNFT, the mean and variance of the phase $\theta(\tau)=\int_{0}^{\tau}\dot{\theta}dt$
can be calculated as $\left\langle \theta(t)\right\rangle \simeq\omega_{s}t$
and $\left\langle (\theta(t)-\left\langle \theta(t)\right\rangle )^{2}\right\rangle \approx\varepsilon^{2}t/Vr_{m}^{2}$.
Hence the velocity $v_{\theta}$ is simply $\omega_{s}$ and the phase
diffusion constant is given by $D_{\theta}\simeq\varepsilon^{2}/2Vr_{m}^{2}.$
It is also possible to obtain the theoretical expression for $\dot{W}$
by using the SNFT, which is after some manipulation given by $\dot{W}\simeq\left(L_{12}-L_{21}\right)V\omega_{s}^{2}r_{m}^{2}$,
where $L_{12}$ and $L_{21}$ are model-dependent parameters determined
by the linear transformation of $F\left(\bm{x}\right)$ at the fixed
point $\bm{x}_{s}$ (see Appendix A for more details), and being independent
of the control parameter $\alpha$ and system size. Consequently,
the transport efficiency reads
\begin{equation}
\eta_{\theta}\simeq\frac{\omega_{s}}{\varepsilon^{2}\left(L_{12}-L_{21}\right)}=\frac{\omega+\alpha\left|C_{i}/C_{r}\right|}{\varepsilon^{2}\left(L_{12}-L_{21}\right)}\label{eq:eta_theta}
\end{equation}
and the TUR asserts that $\eta_{\theta}\le1$. Although the expression
of $\eta_{\theta}$, Eq.(\ref{eq:eta_theta}), gives no hint that
the TUR holds, we indeed demonstrate numerically in our previous work
that for the well-known Brusselator model \cite{cao2020design}, $\eta_{\theta}\sim0.4$
which is far below the upper bound 1.0 in the vicinity of the Hopf
bifurcation.

For oscillation systems, another important quantity is the phase sensitivity
quantifying the ability of the biochemical circuits to respond to
external signal \citep{hasegawa2014optimal,hasegawa2014circadian}.
Instead of dealing with the entire system, we employ the phase reduction
method \citep{kuramoto2003chemical,goldobin2010dynamics} to reduce
the whole state space to a single phase variable $\phi$ characterizing
the timing of oscillation, and the phase sensitivity $\kappa$ can
be obtained by comparing the phase shift after perturbations. The
phase $\phi$ in Eq.(\ref{eq:Phi}) is defined on the limit cycle
of the unperturbed oscillations, and the definition can be expanded
into the entire $\bm{x}$-space by introducing the isochron (the two
states are assigned the same phase if trajectories originated from
two states converge onto the limit cycle at the same time). Following
this definition, the deterministic phase evolution equation can be
expressed as $\dot{\phi}=\Omega=\nabla_{\bm{x}}\phi\cdot\bm{F}(\bm{x})$.
For a weak external signal $\bm{\beta}(t)$, the deterministic term
reads $\bm{F}_{\alpha}(\bm{x})=\bm{F}(\bm{x})+k\bm{\beta}(t)$ with
$k$ the control parameter, and the phase shift incurred by a parametric
perturbation $k\to k+\delta k$ can be obtained as $\dot{\phi}=\Omega+\delta k\left[\nabla_{\bm{x}}\phi\cdot\bm{\beta}(t)\right]$.
Then, the global phase sensitivity parameter $\kappa$ can be defined
as the normalized value of signal-independent factor $\nabla_{\bm{x}}\phi$
along the limit cycle with $r=r_{m}$. For oscillations near the Hopf
bifurcation with $\sqrt{-2C_{r}\varepsilon^{2}/V}<\left|\alpha\right|\ll\left|C_{r}/C_{i}\right|$,
the phase sensitivity $\kappa$ can be approximately calculated as
$\kappa\approx\partial\omega_{s}/\partial\alpha$ \citep{fei2018design,cao2020design},
i.e. (see Eq.\ref{eq:Omega_S})

\begin{equation}
\kappa=\begin{cases}
\frac{\left|C_{i}\right|\varepsilon^{2}}{2\alpha^{2}V} & \alpha<0\\
\left|\frac{C_{i}}{C_{r}}\right| & \alpha>0
\end{cases}.\label{eq:kappa}
\end{equation}

\subsection{Pearson Correlation Coefficient}

Here, we investigate the formulation of a scheme for the characterization
of correlations between the oscillatory amplitude and phase based
on a statistical measure known as the Pearson correlation coefficient,
which has been commonly used in the context of quantum entanglement
\citep{maccone2015complementarity,pozsgay2017covariance,jebarathinam2020pearson}
and filtering theorem \citep{benesty2009pearson,benesty2008importance}.
The Pearson correlation coefficient for any two random variables $R$
and $Q$ is defined as $\chi(R,Q)=\text{Cov}(R,Q)/\sqrt{\text{Var}(R)\text{Var}(Q)}$
with $\text{Cov}(R,Q)=\left\langle RQ\right\rangle -\left\langle R\right\rangle \left\langle Q\right\rangle $
the covariance. The values of Pearson correlation coefficient lie
between $-1$ and $1$.

Then, we start to calculate the Pearson correlation coefficient $\chi_{n}^{2}=\chi^{2}(r^{n},\theta)$
between the two observables $R(\tau)=\theta(\tau)=\int_{0}^{\tau}\dot{\theta}(t)dt$
and $Q_{r,n}=\int_{0}^{\tau}r^{n}\left(t\right)dt$, where the exponent
$n$ quantifies the order of correlation between the oscillatory amplitude
and phase. By using the SNFE, we find that the change rate of the
covariance, $\lim_{\tau\to\infty}\frac{1}{\tau}\text{Cov}(r^{n},\theta;\tau)=\left\langle \left\langle r^{n}\dot{\theta}\right\rangle \right\rangle _{ss}-\left\langle \left\langle \dot{\theta}\right\rangle \right\rangle _{ss}\left\langle \left\langle r^{n}\right\rangle \right\rangle _{ss}$,
between oscillatory phase and amplitude is related to the higher-order
moment of the amplitude as

\begin{align}
\lim_{\tau\to\infty}\frac{1}{\tau}\text{Cov}(r^{n},\theta;\tau) & =\frac{1}{2\pi}\int_{0}^{2\pi}d\theta\int_{0}^{\infty}drr^{n}\dot{\theta}P_{ss}(r)\nonumber \\
 & \hphantom{}\hphantom{}-(\omega+C_{i}\left\langle r^{2}\right\rangle _{ss})\left\langle r^{n}\right\rangle _{ss}\nonumber \\
 & \approx C_{i}\left(\left\langle r^{n+2}\right\rangle _{ss}-\left\langle r^{n}\right\rangle _{ss}\left\langle r^{2}\right\rangle _{ss}\right).\label{eq:Cov}
\end{align}

Particularly, we choose $n=2$ to calculate the covariance between
oscillatory phase and amplitude. According to Eq.(\ref{eq:Phi}),
the change rate of covariance is (see Appendix B for detailed derivation)
\begin{equation}
\lim_{t\to\infty}\frac{1}{\tau}\text{Cov}(r^{2},\theta;\tau)=\begin{cases}
-\frac{2C_{i}C_{r}r_{m}^{4}\varepsilon^{2}}{\alpha^{2}V} & \alpha>0\\
0 & \alpha<0
\end{cases}.\label{eq:Cov-1}
\end{equation}
Here, we need to emphasize that our theoretical expression for normal
oscillations ($\alpha>0$) holds in the region near the Hopf bifurcation
where $\sqrt{-2C_{r}\varepsilon^{2}/V}<\alpha\ll\left|C_{r}/C_{i}\right|$.
It can be found that the phase and amplitude are highly decoupled
with the covariance $\lim_{t\to\infty}\text{Cov}(r^{n},\theta;\tau)/\tau\approx0$
in the subcritical region ($\alpha<0$). The highly decoupling feature
is also the reason why the sensitivity for noise-induced oscillations
($\kappa\sim V^{\delta},\delta=-1$) is typically smaller than the
normal oscillations ($\kappa\sim V^{\delta},\delta=0$), i.e., the
oscillatory amplitude's adaptation to phase shift incurred by perturbation
is much slower in the subcritical region.

Then, we start to calculate the Pearson correlation coefficient $\chi_{2}^{2}=\chi^{2}(r^{2},\theta)$,
which reads as

\begin{align}
\chi_{2}^{2} & =\frac{\left[\text{Cov}(r^{2},\theta)\right]^{2}}{\text{Var}(r^{2})\text{Var}(\theta)}\nonumber \\
 & \approx\frac{C_{i}^{2}\left|\left\langle r^{4}\right\rangle _{ss}-\left\langle r^{2}\right\rangle _{ss}^{2}\right|}{D_{\theta}}.\label{eq:PCC-1}
\end{align}
From Eqs. (\ref{eq:Cov-1}) and (\ref{eq:PCC-1}), the Pearson correlation
coefficient $\chi_{2}^{2}$ for normal oscillations ($\alpha>0$)
can be obtained as

\begin{equation}
\chi_{2}^{2}=\begin{cases}
2\alpha\left(\frac{C_{i}}{C_{r}}\right)^{2} & \alpha>0\\
0 & \alpha<0
\end{cases}.\label{eq:Cov-1-1}
\end{equation}
which is independent of the system size $V$. The Pearson correlation
coefficient $\chi_{n}^{2}=\chi^{2}(r^{n},\theta)$ for $n\neq2$ can
be calculated numerically.

\subsection{Improved TUR}

Recently, it was proposed by Dechant and Sasa that increasing the
number of observables will achieve tighter bounds than the conventional
TUR \citep{dechant2021improving,dechant2018multidimensional}. To
be precise, they defined a generalized transport efficiency as $\eta(R,Q)=\eta(R)+\chi^{2}(R,Q)$,
where $Q=\int_{0}^{\tau}dtq(\bm{x},t)$ is the time-integral of a
state-dependent (non-current) observable $q(\bm{x},t)$, and $\chi(R,Q)$
is the Pearson correlation coefficient between $Q$ and the current
observable $R$. Interestingly, they found that the generalized transport
efficiency, $\eta(R,Q)$, is also smaller than $1$ just like the
conventional one, $\eta(R)$. Therefore, the generalized transport
efficiency $\eta(R,Q)\ge\eta(R)$ provides an improved estimator for
energy dissipation than the conventional one,
\begin{align}
\Delta W_{\text{TUR}} & =\frac{2k_{B}T\left\langle R\right\rangle ^{2}}{\text{Var}(R)}\nonumber \\
 & \le\Delta W_{\text{I}}=\frac{2k_{B}T\left\langle R\right\rangle ^{2}}{\text{Var}(R)[1-\chi^{2}(R,Q)]}\nonumber \\
 & \le\Delta W.\label{eq:iTUR}
\end{align}
It can be found that how much the estimation can be improved is directly
related to the value of Pearson correlation coefficient between the
chosen observables, and the two observables we chose above, the oscillatory
amplitude and oscillatory phase, meet the conditions of use.

Based on Eq.(\ref{eq:Cov-1-1}) and (\ref{eq:iTUR}), the explicit
expression for the generalized transport efficiency can be obtained
as $\eta(r^{2},\theta)=\eta(\theta)+2\alpha(C_{i}/C_{r})^{2}\le1$
when $\alpha>0$. Therefore, we eventually get an efficiency-sensitivity
trade-off relation for normal oscillations,
\begin{equation}
\eta(\theta)+2\alpha\kappa^{2}=\frac{v_{\theta}^{2}}{D_{\theta}\dot{W}}+2\alpha\kappa^{2}\le1,\label{eq:tra}
\end{equation}
which is the main result of our paper, showing that both phase accuracy
$D_{\theta}^{-1}$ and phase sensitivity $\kappa$ can be improved
simultaneously only by increasing the energy dissipation rate $\dot{W}$
without sacrificing the phase speed $v_{\theta}$ \citep{hasegawa2019uncertainty}.
More importantly, such trade-off relation provides an improved estimator
for the dissipation rate,

\begin{equation}
\Delta W_{\text{I,2}}=\frac{2k_{B}T\left\langle \theta\right\rangle ^{2}}{\text{Var}(\theta)(1-\chi_{2}^{2})},
\end{equation}
than the conventional TUR, and the improvement of it is
\begin{equation}
\frac{\Delta W_{\text{I,2}}}{\Delta W_{\text{TUR}}}=\frac{\dot{W}_{\text{I,2}}}{\dot{W}_{\text{TUR}}}=\frac{1}{1-2\alpha\kappa^{2}}>1
\end{equation}
with the TUR estimator $\dot{W}_{\text{TUR}}=\lim_{\tau\to\infty}\Delta W_{\text{TUR}}/\tau$
and the improved estimator $\dot{W}_{\text{I,2}}=\lim_{\tau\to\infty}\Delta W_{\text{I,2}}/\tau$.

Several conclusions can be obtained as follows. Firstly, according
to the trade-off relation Eq.(\ref{eq:tra}), it can be found that
the precision of dissipation inference will be further improved by
enhancing the phase sensitivity of biochemical oscillations. In actual
experimental design, a feasible strategy to achieve a higher phase
sensitivity of the networks is to enhance the phase-amplitude coupling
strength $C_{i}$ by maximizing the net flux of the phase-advancing
pathway relative to that of the phase-retreating pathway \citep{fei2018design}.
Thus, we believe that our analyses provide realizable guidelines for
improving the precision of dissipation estimation for biochemical
oscillations.

Secondly, we find that the generalized transport efficiency $\eta(r^{2},\theta)=\eta(\theta)+2\alpha(C_{i}/C_{r})^{2}$
is independent of the system size $V$. Since the magnitude of the
internal noise is proportional to $V^{-1/2}$, it can be revealed
that our improved scheme is not negatively affected by the internal
noise in the system.

Thirdly, since the phase and amplitude are highly decoupled in subcritical
region ($\alpha<0$), such scheme cannot be applied to improve the
estimation of the energy dissipation for noise-induced oscillations.

In the following, we further highlight our motivation. As stated above,
it has been proposed that the TUR provides a powerful tool to estimate
energy dissipation. Recently, this bound has been
optimized to provide a more accurate estimation \cite{li2019quantifying,manikandan2020inferring,otsubo2020estimating,van2020entropy,busiello2019hyperaccurate,manikandan2018exact,gingrich2017inferring,kim2020learning,busiello2022hyperaccurate}
and even realize equality \cite{dechant2021continuous,manikandan2020inferring,otsubo2020estimating,van2020entropy},
which is of great significance. Particularly, Manikandan $et$ $al.$
have found that the TUR estimates entropy production exactly in the
very short time limit, if the observed current is optimally chosen,
which provides a powerful strategy for the dissipation inference \cite{manikandan2020inferring}.
Some optimization procedure needs to be utilized to obtain the optimal
current, where the similar manipulations have been used to get the
hyperaccurate currents \cite{li2019quantifying,busiello2019hyperaccurate,busiello2022hyperaccurate}.
Other techniques such as the gradient ascent in machine learning have
also been applied to construct the short-time limit TUR estimator
by Otsubu $et$ $al.$ \cite{otsubo2020estimating} and Vu $et$ $al.$
\cite{van2020entropy}. However, the related procedure may be difficult
to follow than measuring the dissipation itself \cite{dechant2021improving}.
To be specific, in Ref \cite{manikandan2020inferring}, Manikandan
$et$ $al.$ tested their inference scheme by numerically calculating
the optimal current rely on linear combinations of the basis. In Ref
\cite{otsubo2020estimating}, Otsubu $et$ $al.$ demonstrated that
their learning protocol performs well by numerical experiments in
nonlinear Langevin dynamics. In addition to the TUR-based approach,
Frishman and Ronceray proposed a principled method, stochastic force
inference, to evaluate the corresponding entropy production based
on approximating force fields and diffusion coefficients \cite{frishman2020learning}.
To sum up, all the dissipation estimations mentioned above require
some specific preprocessing, thus increasing the statistical effort.
As a comparison, only the oscillatory phase and amplitude need to
be tracked by using the scheme proposed by us, which is readily accessible,
showing its benefit for experimentally application. Also, the results
of our manuscript have demonstrated that the estimation of energy
dissipation is considerably improved by considering the correlations
between observables, no matter how far from equilibrium the system
is. In addition, since biochemical functional benefits from operating
at the edge of instability, studying the oscillatory behavior near
Hopf bifurcation points can bring general inspiration \cite{munoz2018colloquium}.
Therefore, we believe that our formulations provide an efficient estimator
in terms of experimentally accessible quantities.

\begin{figure}
\begin{centering}
\includegraphics[width=1\columnwidth]{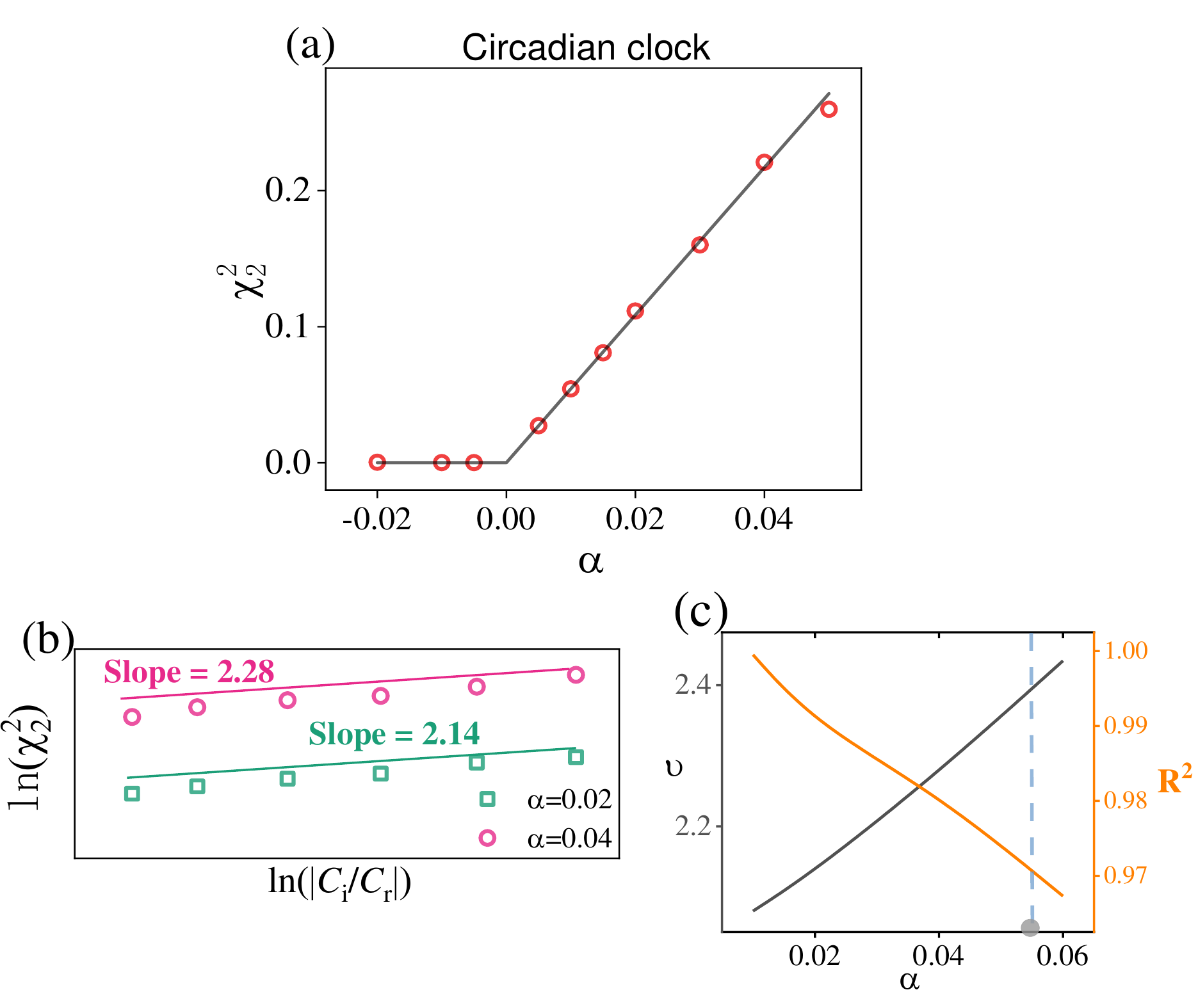}
\par\end{centering}
\caption{(a) Pearson correlations $\chi_{2}^{2}$ as a function of the control
parameter $\alpha$ for the circadian clock model. The value of $\chi_{2}^{2}$
changes sharply near the critical point $\alpha=0$, due to the bifurcation
phenomenon. Line: theory. Dots: simulation. (b) The Pearson correlations
$\chi_{2}^{2}$ as a function of the phase sensitivity $\kappa=\left|C_{i}/C_{r}\right|$
(green and pink dots). The slopes $\nu$ for $\chi_{2}^{2}\propto\kappa^{\nu}$
have been calculated from fitting the numerical data. (c) The slope
$\nu$ and goodness $R^{2}$ for the linear fit between $\ln(\chi_{2}^{2})$
and $\ln\kappa$ as a function of the control parameter $\alpha$.
The gray circle and blue dotted line represent the range in which
the scaling behavior $\chi_{2}^{2}\propto\kappa^{2}$ holds. The establishment
of the scaling behavior reveals how far from the Hopf bifurcation
the trade-off relation between dissipation and phase sensitivity satisfies.
The system size $V=1.6\times10^{5}$.}

\label{fig:1}
\end{figure}

\section{Simulations}

\begin{figure}
\begin{centering}
\includegraphics[width=0.8\columnwidth]{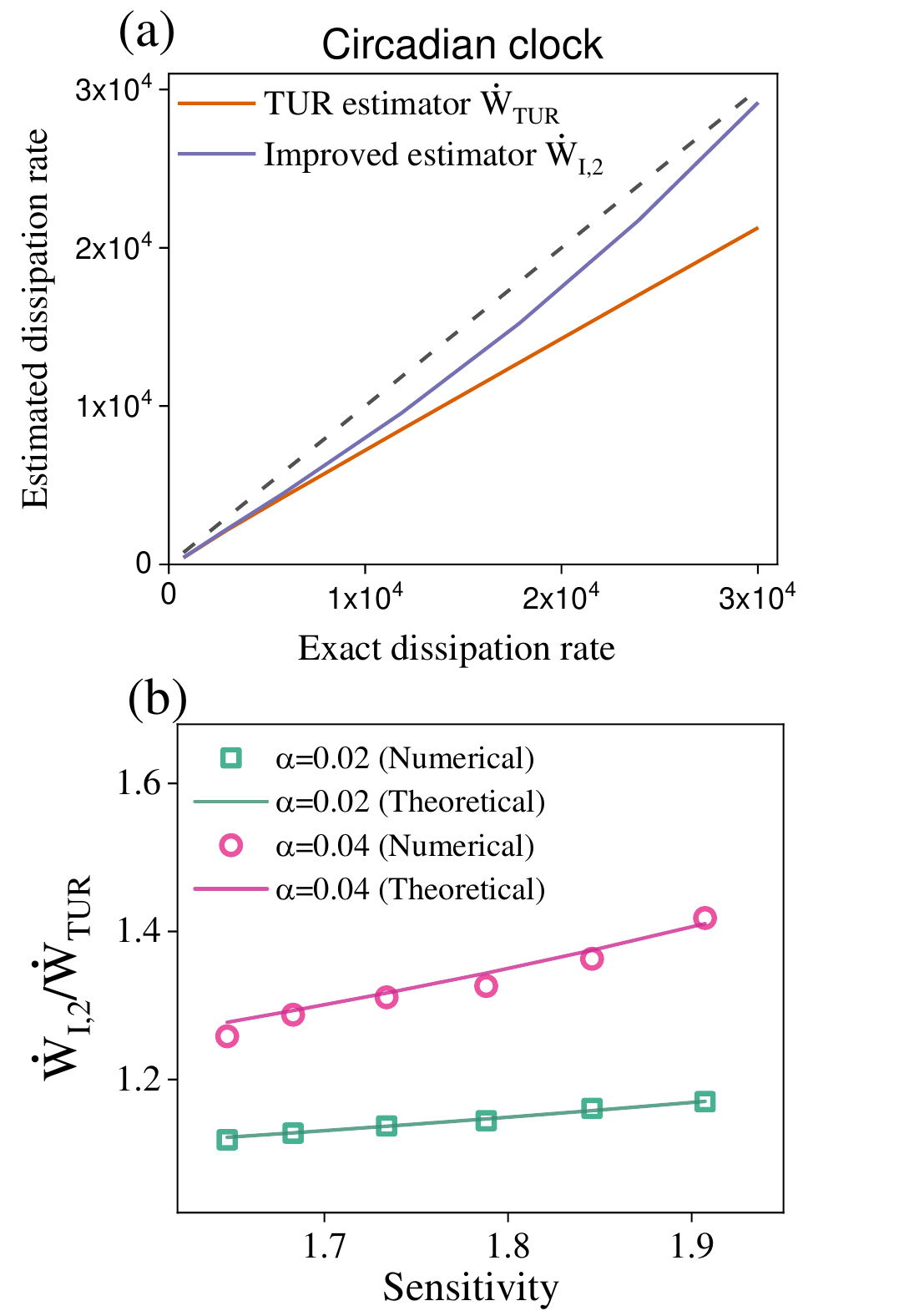}
\par\end{centering}
\caption{(a) Estimations of the dissipation rate as a function of the exact
dissipation rate $\dot{W}$ for $\alpha>0$. It can be observed that
the improved estimator $\dot{W}_{\text{I,2}}$ outperforms the TUR
estimator $\dot{W}_{\text{TUR}}$. The parameters of circadian clock
model can be found in Appendix C. (b) The improvement of the tighter
bound $\dot{W}_{\text{I,2}}/\dot{W}_{\text{TUR}}$ as a function of
the phase sensitivity $\kappa=\left|C_{i}/C_{r}\right|$. The numerical
results (dots) verify our analytical expressions (lines). The system
size $V=1.6\times10^{5}$.}

\label{fig:2}
\end{figure}

In this section, we illustrate the formal analytical results of the
above section within numerical simulations of the circadian clock
model \citep{hou2003internal}, describing how living systems keep
an internal sense of time. The circadian clock model considered here
incorporates the transcription of the gene (G) involved in the biochemical
clock and transport of the mRNA (R) into the cytosol where it is translated
into clock proteins ($\text{P}_{C}$) and degraded. The protein can
be degraded or transported into the nucleus ($\text{P}_{N}$) where
it exerts a negative regulation on the expression of its gene. For
the parameters we examine (see Appendix C for details), the Hopf bifurcation
point locates at $v_{s}\simeq0.25725$ with $v_{s}$ the transcription
rate of mRNA. In addition, parameter values used in the stochastic
normal form theory can be calculated from simulations as $C_{r}\simeq-0.3474$,
$C_{i}\simeq0.5722$ and $\varepsilon^{2}\simeq0.3556$. By adjusting
the transition rates, the values of $C_{i}$ and $C_{r}$ will change,
and can also be obtained.

By using the Euler methods, we numerically calculate Eqs.(\ref{eq:dr})
and (\ref{eq:dth}) with a time step of $0.002$. Generally, after
a long time $t_{ss}=10^{5}$ to ensure the system reaches the steady
state, $2\times10^{5}$ trajectories with the length $t_{0}=1$ are
used to get the Pearson correlation coefficient, $\chi_{n}^{2}=\chi^{2}(r^{n},\theta)=\left[\text{Cov}(r^{n},\theta)\right]^{2}/\text{Var}(r^{n})\text{Var}(\theta)$
for $Q_{r,n}=\int_{0}^{\tau}r^{n}\left(t\right)dt$ with $n>0$ the
power, which yields the corresponding improved estimator as
\[
\Delta W_{\text{I,n}}=\frac{2k_{B}T\left\langle \theta\right\rangle ^{2}}{\text{Var}(\theta)(1-\chi_{n}^{2})}.
\]
Then, the TUR estimator $\dot{W}_{\text{TUR}}=\lim_{\tau\to\infty}\Delta W_{\text{TUR}}/\tau$
and the improved estimator $\dot{W}_{\text{I,n}}=\lim_{\tau\to\infty}\Delta W_{\text{I,n}}/\tau$
can be obtained numerically. The corresponding improvement reads as
\[
\frac{\dot{W}_{\text{I,n}}}{\dot{W}_{\text{TUR}}}=\frac{1}{1-\chi^{2}(r^{n},\theta)}>1.
\]
On the other hand, the exact dissipation rate $\dot{W}$ is obtained
from the simulation data of Eq.(\ref{eq:CLE}) (see Appendix A for
details).

In Fig.\ref{fig:1}(a), the dependence of the Pearson correlations
$\chi_{n}^{2}$ ($n=1,2,3$) on the control parameter $\alpha$ are
depicted for the circadian clock model. The value range of the control
parameter $\alpha$ ensures the establishment of the SNFT. For noise-induced
oscillations in the subcritical region($\alpha<0$), the Pearson correlation
coefficients are almost zero, and they increase significantly after
the control parameter crossing the critical point $\alpha=0$ to reach
the supercritical region for normal oscillations ($\alpha>0$). Those
results verify our prediction that the correlation between oscillatory
phase and amplitude is highly decoupled and not sufficient to improve
the estimation of energy dissipation for noise-induced oscillations.
In addition, we notice that numerical results (dots) of the Pearson
correlations are in good agreement with our theoretical predictions,
Eq.(\ref{eq:Cov-1-1}) (line). Further, in Fig.\ref{fig:1}(b), we
plot the Pearson correlations $\chi_{2}^{2}$ as a function of the
phase sensitivity $\kappa=\left|C_{i}/C_{r}\right|$. The scaling
behaviors are consistent with our analytical result $\chi_{2}^{2}\propto\kappa^{2}$,
further confirming our theory. It can be found that the slope $\nu$
for $\chi_{2}^{2}\propto\kappa^{\nu}$ is closer to the analytical
prediction $\nu=2$ for smaller $\alpha$, showing that our theory
is more accurate for near Hopf bifurcation region. To further explore
the extent to which our formulation holds generally away from a Hopf
Bifurcation, we plot the slope $\nu$ and goodness $R^{2}$ for the
linear fit between $\ln(\chi_{2}^{2})$ and $\ln\kappa$ as a function
of the control parameter $\alpha$ in Fig.\ref{fig:1}(c). The gray
circle and blue dotted line represent the range in which the scaling
behavior $\chi_{2}^{2}\propto\kappa^{2}$ holds ($R^{2}\ge0.97$).
The establishment of the scaling behavior reveals how far from the
Hopf bifurcation the trade-off relation between dissipation and phase
sensitivity satisfies. For the circadian clock model, our formulation
holds for $\alpha\le0.055$.

In Fig.\ref{fig:2}(a), we show both the TUR estimator $\dot{W}_{\text{TUR}}$
and the improved estimator $\dot{W}_{\text{I}}$ for the circadian
clock model to demonstrate how much the estimation of energy dissipation
can be improved. The conventional TUR, while a commonly used dissipation
estimator, only provides a trivial bound with the 0.6 efficiency,
and the improved estimator $\dot{W}_{\text{I}}$ is much closer to
the exact value $\dot{W}$. On the other hand, it can be found that
$\dot{W}\ge\dot{W}_{\text{I}}$, which verifies the efficiency-sensitivity
trade-off relation we proposed {[}Eq.(\ref{eq:tra}){]}. In Fig.\ref{fig:2}(b),
we show the relationship between the improvement of the tighter bound
$\dot{W}_{\text{I}}/\dot{W}_{\text{TUR}}$ and phase sensitivity $\kappa=\left|C_{i}/C_{r}\right|$.
The theoretical predictions $\dot{W}_{\text{I}}/\dot{W}_{\text{TUR}}=\frac{1}{1-2\alpha(C_{i}/C_{r})^{2}}$
are in good agreement with the numerical results, which demonstrates
that the estimation of the dissipation can be improved by enhancing
the phase sensitivity of biochemical oscillations.

\begin{figure}
\begin{centering}
\includegraphics[width=1\columnwidth]{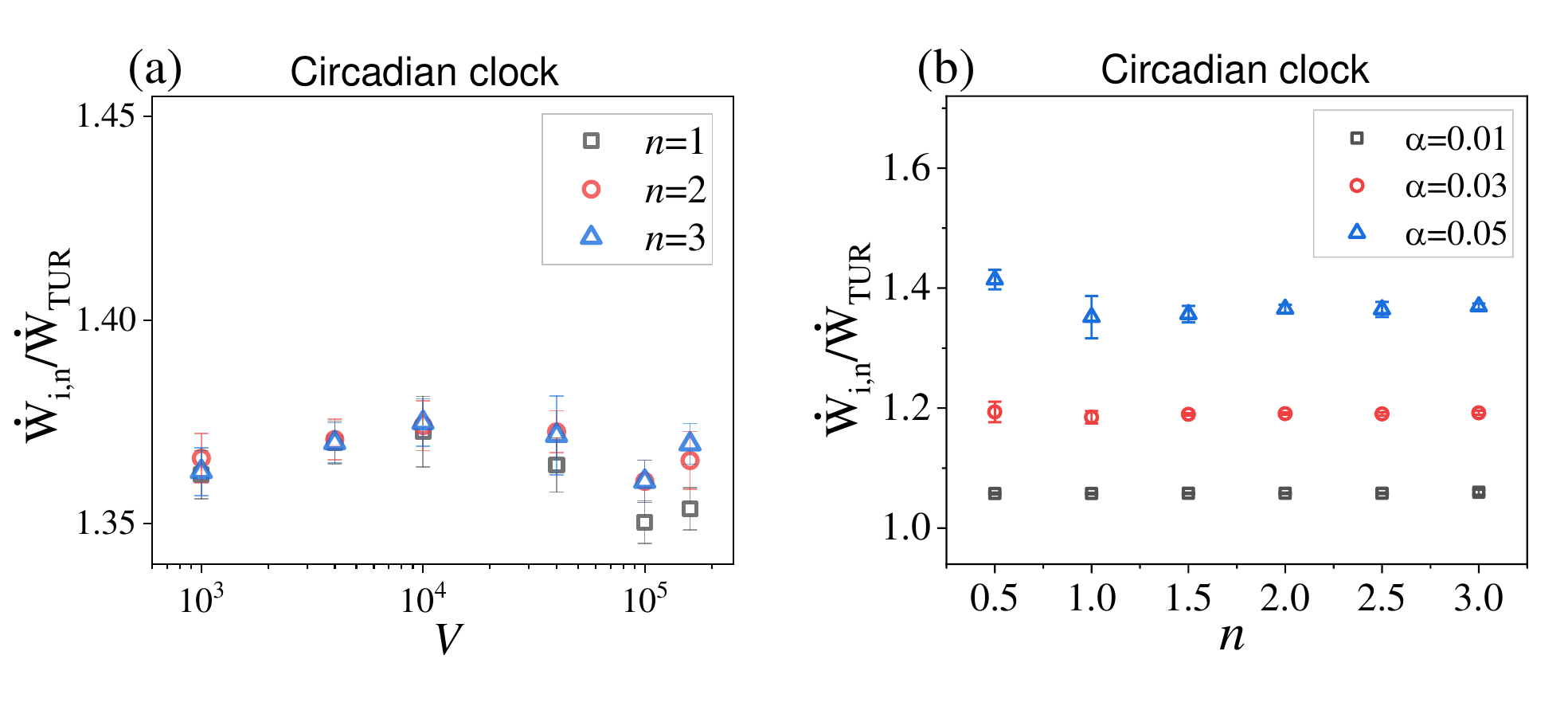}
\par\end{centering}
\caption{(a) The improvements of the dissipation estimation $\dot{W}_{\text{I}}/\dot{W}_{\text{TUR}}$
($n=1,2,3$) as a function of the system size $V$ for the circadian
clock model in normal oscillation region ($\alpha>0$). The control
parameter $\alpha=0.05$. (b) The improvements of the dissipation
estimation $\dot{W}_{\text{I}}/\dot{W}_{\text{TUR}}$ as a function
of the power of the amplitude, $n$, for the circadian clock model.
The values are independent of the power. The error bars represent
the standard deviation obtained from independent trials. The system
size $V=1.6\times10^{5}$.}

\label{fig:3}
\end{figure}

In Fig.\ref{fig:3} (a), we have shown that the improvements of the
dissipation estimation $\dot{W}_{\text{I,n}}/\dot{W}_{\text{TUR}}$
change little with the system size $V$ in normal oscillations ($\alpha>0$).
Moreover, we numerically test whether the power of amplitude observables
$n$ affect the improvement $\dot{W}_{\text{I,n}}/\dot{W}_{\text{TUR}}$
in details. In Fig.\ref{fig:3} (b), it can be observed that values
of $\dot{W}_{\text{I,n}}/\dot{W}_{\text{TUR}}$ change little for
different choices of the power of the amplitude observable $Q_{r,n}(\tau)=\int_{0}^{\tau}r^{n}(t)dt$.

To further demonstrate the broad application of the proposed improved
estimation, we have also applied them to another well-known biochemical
oscillation system, the Brusselator model. Other details of the model
and parameters can be found in Appendix C. As shown in Fig.\ref{fig:4},
we find that our main results, such as the theoretical expression
of Pearson correlations {[}Eq. (\ref{eq:Cov-1-1}){]} and the improved
estimation {[}obtained from Eq. (\ref{eq:iTUR}){]}, still hold in
the Brusselator model. In Fig.\ref{fig:4}(a), it can be observed
that numerical results (dots) of the Pearson correlations are in good
agreement with our theoretical predictions, Eq.(\ref{eq:Cov-1-1})
(line). In Fig.\ref{fig:4}(b), both the TUR estimator $\dot{W}_{\text{TUR}}$
and the improved estimator $\dot{W}_{\text{I}}$ has been depicted
for the Brusselator model. As expected, the conventional TUR estimator
only yields a loose bound, and our improved estimator $\dot{W}_{\text{I}}$
is much more accurate.

\begin{figure}
\begin{centering}
\includegraphics[width=1\columnwidth]{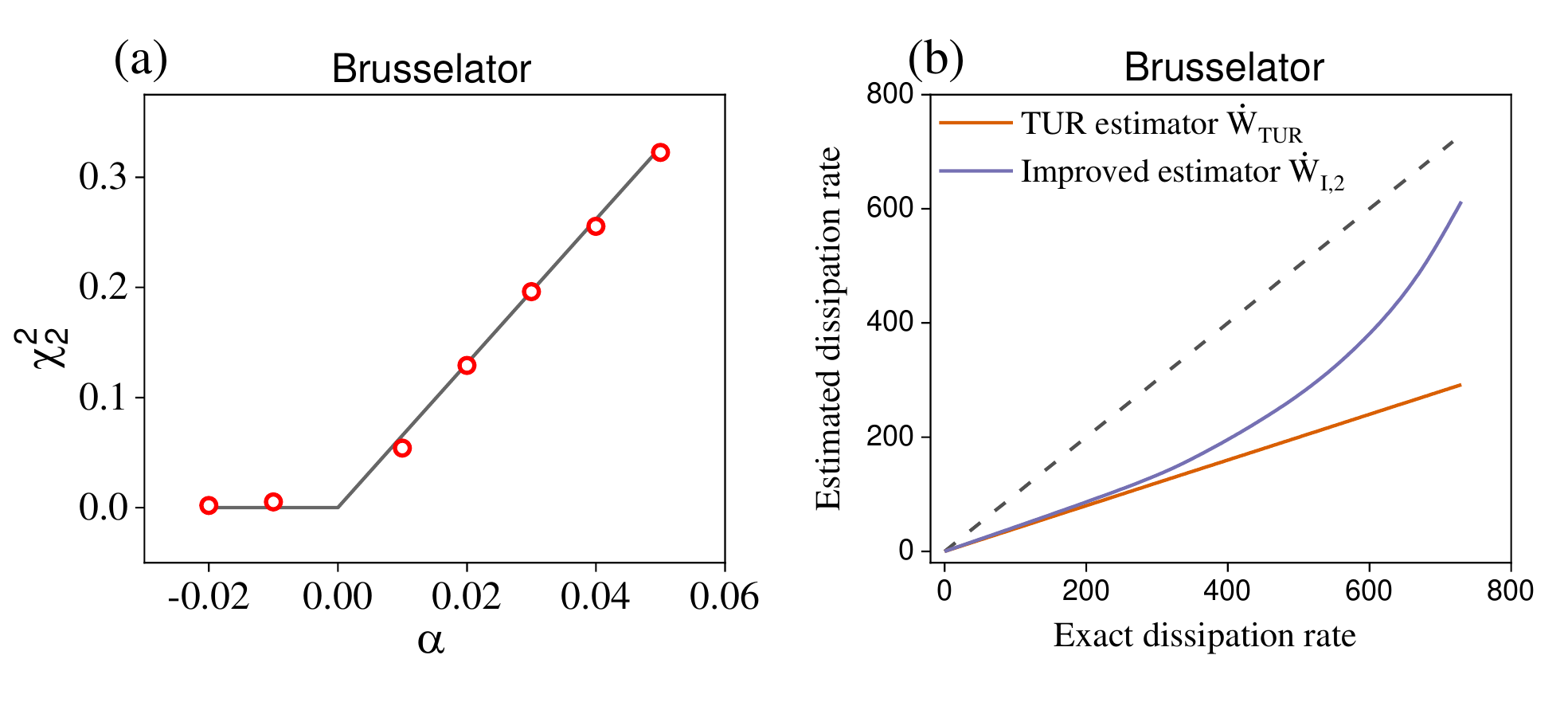}
\par\end{centering}
\caption{(a) Pearson correlations $\chi_{2}^{2}$ as a function of the control
parameter $\alpha$ for the Brusselator model. The value of $\chi_{2}^{2}$
changes sharply near the critical point $\alpha=0$, due to the bifurcation
phenomenon. Line: theory. Dots: simulation. (b) Estimations of the
dissipation rate as a function of the exact dissipation rate for $\alpha>0$.
It can be observed that the improved estimator $\dot{W}_{\text{I,2}}$
outperforms the TUR estimator $\dot{W}_{\text{TUR}}$. The parameters
of the Brusselator model can be found in Appendix C. The system size
$V=1.6\times10^{5}$.}

\label{fig:4}
\end{figure}

\section{Discussion}

In this paper, we proposed an improved estimation for the energy dissipation
of biochemical oscillations by using the Pearson correlations between
oscillatory phase and amplitude, which are easily accessible in experimental
observations. Both the analytical and numerical results demonstrate
that such scheme can be further improved by enhancing the phase sensitivity
of systems. In addition, it has been revealed by us that the validity
of our scheme is independent of the system size and the power of oscillatory
amplitude.

In our previous work \citep{cao2020design}, we have found that the
dissipation rate $\dot{W}\sim V^{\gamma}$, with $\gamma=1$ for supercritical
region ($\alpha>0$), $\gamma=1/2$ for the critical point ($\alpha=0$)
and $\gamma=0$ for subcritical region ($\alpha<0$), showing that
biochemical oscillations have a much lower energy dissipation for
noise-induced oscillation. Intuitively, one might think that less
dissipation will lead to a easier estimation, however, the estimator
$\Delta W_{\text{I,n}}$ introduced by us is not applicable for improving
the estimation of dissipation due to the highly decoupling of the
phase and amplitude, which is deserved for further study.

Biomolecules, especially proteins, can act as tiny and highly functional
machines, such as kinesin \cite{verhey2011kinesin} and ribosome \cite{aitken2010single}.
To probe the operation of these bimolecular machines, it is not enough
just to know their structure, one needs to understand how the structure
generates specific conformational dynamics. Meanwhile, how much energy
the biological machine dissipates to perform certain functions is
also a major issue \cite{lan2012energy}. Particularly, molecular
dynamics (MD) simulation is a primary technique for studying bimolecular
machines, producing information about the conformational dynamics
with spatial and temporal resolutions. For the molecular systems,
the key ingredients of the slow kinetics can be obtained by using
SNFT based on stochastic averaging \cite{cao2020design} or variational
approach with MD simulations \cite{nuske2014variational}(Perhaps
some short time dynamics with energy input). Then, the findings of
this manuscript that correlations between observables can yield improved
estimation could be applied in these systems based on the accurate
capture of the dominant motion. However, it is still important to
note that fast processes may also have finite correction to the thermodynamic
quantities \cite{seifert2019stochastic}. As stochastic normal form
equations can be extended to other oscillatory systems related to
other types of bifurcations, such as relaxation oscillations, we believe
that our scheme may have a wider range of applications.
\begin{acknowledgments}
This work is supported by MOST(2018YFA0208702), NSFC (32090044, 21790350,
21521001).
\end{acknowledgments}

\section*{Author declarations}

The authors have no conflicts to disclose.

\section*{Data availability}

The data that support the findings of this study are available from
the corresponding author upon reasonable request.

\appendix



\section{The derivation of normal form and calculation of transport efficiency}

In this section, we introduce the derivation of stochastic normal
form equation and the calculation of conventional transport efficiency
for self-consistency.

\subsection{Stochastic normal form theory}

Firstly, we assume that the deterministic form of the chemical Langevin
equation, Eq.(\ref{eq:CLE}), has a unique stable point $\bm{x}_{s}$
with $\bm{F}(\bm{x}_{s})\equiv0$, which loses stability at the supercritical
HB $\mu=\mu_{c}$, where $\mu$ is the control parameter. Based on
the Hopf theorem \citep{hassard1981theory}, the Jacobian matrix $\bm{J}$,
whose components $J_{ij}=\left(\partial f_{i}/\partial x_{j}\right)|_{\bm{x}=\bm{x}_{s}}$,
has a pair of conjugate eigenvalues $\lambda_{\pm}=\alpha(\mu)\pm i\omega$
with $\alpha(\mu_{c})=0$. The other $N-2$ eigenvalues of $\bm{J}$,
$-\lambda_{j(\ge3)}$, all have negative real parts with absolute
values considerably larger than $0$. Performing the variable transformation
$\bm{u}=\bm{T}^{-1}(\bm{x}-\bm{x}_{s})$, the linear part of Eq.(\ref{eq:CLE})
can be transformed to Jordan form as $\dot{\bm{u}}=\bm{\Lambda\bm{u}}+O(\bm{u}^{2})+\frac{1}{\sqrt{V}}\bm{\eta}\left(t\right),$where
$\bm{\Lambda}=\left(\begin{array}{cc}
\alpha & -\omega\\
\omega & \alpha
\end{array}\right)\oplus\text{diag\ensuremath{\left(-\lambda_{1},\dots,-\lambda_{N}\right)}}$ and $\bm{\eta}=\bm{T}^{-1}\zeta(\bm{x}_{s},t)$ with $\zeta(\bm{x},t)=\sum_{\rho}v_{\rho}^{j}\sqrt{w_{\rho}(\bm{x})}\xi_{\rho}(t)$.
The variances of $\bm{\eta}$ are $\left\langle \eta_{i}(t)\eta_{j}(s)\right\rangle =2D_{ij}\delta(t-s)$
with $\bm{D}=\bm{T}^{-1}\bm{G}(\bm{T}^{-1})^{\text{T}}$. The transformation
is done as follows. Firstly, we calculate the eigenvector $\bm{u}_{+}$
whose eigenvalue $\lambda_{+}=\alpha+i\omega$, and normalize it to
ensure the first non-vanishing component is $1$. Secondly, we construct
a matrix $\bm{T}=(Re\bm{u}_{+},-Im\bm{u}_{+},\bm{r}_{3},\ldots,\bm{r}_{n})$
with $(\bm{r}_{3},\ldots,\bm{r}_{n})$ are any set of real vectors
which span the union of the eigenspaces for $(\lambda_{3},\ldots,\lambda_{n})$.
Finally, it is allowed to perform the change of variables $\bm{x}=\bm{x}_{s}+\bm{T}\bm{u}$.

When the system locates near the HB ($\left|\alpha\right|\ll1$),
the evolution of the oscillatory mode related to $(u_{1},u_{2})$
is much slower than the other $N-2$ stable modes due to the time-scale
separation. Hence, the system's dynamics will be dominated by the
slow motion on a 2D center manifold spanned by the eigenvectors of
$\lambda_{\pm}$. The oscillatory mode are ruled by a normal form
equation involving the time evolution of a complex variable $Z=u_{1}+iu_{2}$,
or a pair of coupled equations for the oscillation amplitude $r$
and phase $\theta$ via $Z=re^{i\theta}$.We follow the standard procedure
to get the normal form,
\begin{align}
\frac{dZ}{dt} & =\left(\alpha+i\omega\right)Z+\left(C_{r}+iC_{i}\right)\left|Z\right|^{2}Z\nonumber \\
 & \hphantom{\hphantom{}}\hphantom{}+\frac{1}{\sqrt{V}}\sum_{\rho}\left(v_{1\rho}^{\prime}+iv_{2\rho}^{\prime}\right)\sqrt{w_{\rho}}\xi_{\rho},
\end{align}
where $\bm{v}_{j\rho}^{\prime}=\left(T^{-1}\bm{v}\right)_{j\rho}$,
i.e.,
\begin{equation}
\frac{dr}{dt}=\left(\alpha r+C_{r}r^{3}\right)+\frac{1}{\sqrt{V}}\sum_{\rho}\chi_{r\rho}\circ\xi_{\rho},
\end{equation}

\begin{equation}
\frac{d\theta}{dt}=\left(\omega+C_{i}r^{2}\right)+\frac{1}{\sqrt{V}}\sum_{\rho}\chi_{\theta\rho}\circ\xi_{\rho}
\end{equation}
with
\begin{equation}
\chi_{r\rho}=\left(v_{1\rho}^{\prime}\cos\theta+v_{2\rho}^{\prime}\sin\theta\right)\sqrt{w_{\rho}},
\end{equation}
\begin{equation}
\chi_{\theta\rho}=\frac{1}{r}\left(-v_{1\rho}^{\prime}\sin\theta+v_{2\rho}^{\prime}\cos\theta\right)\sqrt{w_{\rho}}.
\end{equation}
By using the ``stochastic averaging'' method \citep{arnold1996toward},
the following equation can be obtained
\begin{equation}
\frac{dr}{dt}=\alpha r+C_{r}r^{3}+\frac{K(r)}{V}+\frac{\varepsilon_{r}}{\sqrt{V}}\xi_{r},
\end{equation}
and
\begin{equation}
\frac{d\theta}{dt}=\omega+C_{i}r^{2}+\frac{K(\theta)}{V}+\frac{\varepsilon_{\theta}}{r\sqrt{V}}\xi_{\theta}.
\end{equation}
Here,
\begin{equation}
K(r)=\frac{1}{2\pi}\sum_{\rho}\intop_{0}^{2\pi}d\theta\left(\chi_{r\rho}\partial_{r}\chi_{r\rho}+\chi_{\theta\rho}\partial_{\theta}\chi_{r\rho}\right),
\end{equation}

\begin{equation}
K(\theta)=\frac{1}{2\pi}\sum_{\rho}\intop_{0}^{2\pi}d\theta\left(\chi_{r\rho}\partial_{r}\chi_{\theta\rho}+\chi_{\theta\rho}\partial_{\theta}\chi_{\theta\rho}\right),
\end{equation}
which is related to the coupling effects between amplitude and phase.
$\varepsilon_{r}^{2}=\frac{1}{2\pi}\sum_{\rho}\intop_{0}^{2\pi}d\theta\chi_{r\rho}^{2}$
and $\varepsilon_{\theta}^{2}=\frac{1}{2\pi}\sum_{\rho}\intop_{0}^{2\pi}d\theta\chi_{\theta\rho}^{2}$
are the averaged noise intensities.{} The main purpose of this method
is to approximate the system's dynamics as the Markovian stochastic
process when the system reaches the steady state. Further, by expanding
the reaction rates, $w_{\rho}=\sum_{i+j=0}^{n}w_{\rho}^{ij}(r\cos\theta)^{i}(r\sin\theta)^{j}$,
$K(\theta)$ is zero \citep{hou2006internal}. Thus, the averaged
noise intensities read as
\begin{equation}
\varepsilon_{r}^{2}=\varepsilon_{\theta}^{2}=\frac{1}{2}\sum_{\rho}\left[\left(v_{1\rho}^{\prime}\right)^{2}+\left(v_{2\rho}^{\prime}\right)^{2}\right]w_{\rho}^{00}
\end{equation}
near the Hopf bifurcation point, i.e., the stochastic normal form
equation can be obtained as
\begin{equation}
\dot{r}=\alpha r+C_{r}r^{3}+\frac{\varepsilon^{2}}{2Vr}+\frac{\varepsilon}{\sqrt{V}}\eta_{r}(t),\label{eq:dotr-1}
\end{equation}
\begin{equation}
\dot{\theta}=\omega+C_{i}r^{2}+\frac{\varepsilon}{r\sqrt{V}}\eta_{\theta}(t),\label{eq:dottheta-1}
\end{equation}
where the $i+j\geqslant2$ terms are neglected.

\subsection{Steady state dissipation rate and conventional transport efficiency}

In order to obtain the transport efficiency, we start to calculate
the steady state dissipation rate $\dot{W}$. Based on the framework
of stochastic thermodynamics \citep{seifert2005entropy,seifert2012stochastic,sekimoto2010stochastic,jarzynski2011equalities,gaspard2004fluctuation},
the entropy balance equation reads as $\dot{s}_{tot}(\tau)=\dot{s}_{m}(\tau)+\dot{s}(\tau)$,
where $s_{tot}(\tau)$ is the total entropy production, $s(\tau)$
is the Shannon entropy and $s_{m}(\tau)$ is the entropy flux. As
$s(\tau)=-\ln p\left(\bm{x},\tau\right)$, the change rate of the
Shannon entropy is

\begin{align}
\dot{s}(\tau) & =\left[-\partial_{\tau}p(\bm{x},\tau)+\frac{2V}{p(\bm{x},\tau)}\sum_{i,j}\Gamma_{ij}J_{j}|_{\bm{x}(\tau)}\dot{x}_{i}\right]\nonumber \\
 & \hphantom{}\hphantom{}-V\sum_{i}H_{i}\dot{x}_{i},
\end{align}
where $H_{j}=2\sum_{k}\Gamma_{jk}f_{k}^{\prime}$ ($\bm{\Gamma}=\bm{G}^{-1}$)
with $\widetilde{f}_{k}=f_{k}-1/(2V)\sum_{j}(\partial G_{kj})/(\partial x_{j})$.
Then, the entropy production rate and entropy flux rate can be identified
as $\dot{s}_{tot}(\tau)=-\partial_{\tau}p(\bm{x},\tau)+\frac{2V}{p(\bm{x},\tau)}\sum_{i,j}\Gamma_{ij}J_{j}|_{\bm{x}(\tau)}\dot{x}_{i}$
and $\dot{s}_{m}(\tau)=V\sum_{i}H_{i}\dot{x}_{i}$. As $\dot{s}(\tau)=\lim_{t\to\infty}\left\langle \Delta s\right\rangle /t$
vanishes in the steady state, the averaged entropy production rate
can be obtained as
\begin{equation}
\dot{S}_{tot}=\lim_{t\to\infty}\left\langle \Delta s_{m}\right\rangle /t=V\sum_{i}\left\langle \left\langle H_{i}\dot{x}_{i}\right\rangle \right\rangle _{ss}\label{eq:Stot}
\end{equation}
with $\left\langle \left\langle \cdot\right\rangle \right\rangle _{ss}$
denotes the average over time and steady state \citep{xiao2009stochastic}.

By using the variable transform, the theoretical expression of the
entropy production rate can be calculated in terms of $\bm{u}$, which
reads $\dot{S}_{tot}=2V\left\langle \left\langle \widetilde{\bm{f}}^{\text{T}}\bm{\Gamma}^{\text{T}}\dot{\bm{x}}\right\rangle \right\rangle _{ss}$.
By approximating $\widetilde{\bm{f}}(\bm{x)}\approx\bm{J}\bm{T}\bm{u},$
the entropy production reads

\begin{equation}
\dot{S}_{tot}=2V\left\langle \left\langle \bm{u}^{\text{T}}\bm{L}\dot{\bm{u}}\right\rangle \right\rangle _{ss}=2V\sum_{i,j}L_{ij}h_{ij}
\end{equation}
with $h_{ij}=\left\langle \left\langle u_{i}\dot{u}_{j}\right\rangle \right\rangle _{ss}$.
$\bm{L}=\bm{T}^{\text{T}}\bm{J}^{\text{T}}\bm{\Gamma}^{\text{T}}\bm{T}$
are model-dependent parameters taken the value at the stable point
$\bm{x}_{s}$. Note that in the steady state, $\frac{d}{dt}\left\langle \left\langle u_{i}\dot{u}_{j}\right\rangle \right\rangle _{ss}=0$,
thus we have $h_{ij}=-h_{ji}$. Then, we have that

\begin{align}
h_{12} & =-h_{21}=\left\langle \left\langle r\cos\theta\frac{d}{dt}\left(r\sin\theta\right)\right\rangle \right\rangle _{ss}\nonumber \\
 & =\frac{1}{2\pi}\int_{0}^{2\pi}\dot{\theta}\cos^{2}\theta d\theta\cdot\int_{0}^{\infty}r^{2}p_{ss}\left(r\right)dr\approx\frac{1}{2}\omega_{s}\left\langle r^{2}\right\rangle ,
\end{align}
where the time average is substituted by averaging over $\theta$
due to dominant oscillatory mode. $\omega_{s}=\omega+C_{i}r_{m}^{2}$
is the effective phase angular velocity. Meanwhile, for $j>2,$we
have $h_{1j}=\left\langle \left\langle r\cos\theta\dot{u}_{j}\right\rangle \right\rangle _{ss}\approx0$
and $h_{2j}=\left\langle \left\langle r\sin\theta\dot{u}_{j}\right\rangle \right\rangle _{ss}\approx0$.
For $i,j>2$, one can obtain that $h_{ij}=\left\langle \left\langle u_{i}\dot{u}_{j}\right\rangle \right\rangle _{ss}=(\lambda_{i}-\lambda_{j})D_{ij}/[(\lambda_{i}+\lambda_{j})V]$.
Therefore, the averaged entropy production rate is

\begin{equation}
\dot{S}_{tot}=V\left(L_{12}-L_{21}\right)\omega_{s}\left\langle r^{2}\right\rangle +2\sum_{i,j>2}L_{ij}D_{ij}\frac{\lambda_{i}-\lambda_{j}}{\lambda_{i}+\lambda_{j}}.
\end{equation}
Here, $r_{m}$ is the most probable value of the amplitude in the
steady state with $\partial_{r}P_{ss}\left(r\right)|_{r=r_{m}}=0$.
By going through our derivation, the contributions from the remaining
other $N-2$ stable modes can also be identified as $\dot{S}_{fast}=2\sum_{i,j>2}L_{ij}D_{ij}(\lambda_{i}-\lambda_{j})/(\lambda_{i}+\lambda_{j})$,
which is absent in the expressions obtained by the conventional steady
state formula \cite{cao2015free,fei2018design}. Further, the steady
state dissipation rate (here we set $k_{B}T=1$)

\begin{equation}
\dot{W}=k_{B}T\dot{S}_{tot}\approx V\left(L_{12}-L_{21}\right)\omega_{s}r_{m}^{2}
\end{equation}
Now, we start to calculate the transport efficiency $\eta_{\theta}$.
The mean and variance of the phase $\theta(\tau)=\int_{0}^{\tau}\dot{\theta}dt$
can be calculated as $\left\langle \theta(t)\right\rangle \approx\omega_{s}t$
and $\left\langle (\theta(t)-\left\langle \theta(t)\right\rangle )^{2}\right\rangle \approx\varepsilon^{2}t/Vr_{m}^{2}$,
and the phase diffusion constant is given by $D_{\theta}=\lim_{t\to\infty}\left\langle (\theta(t)-\left\langle \theta(t)\right\rangle )^{2}\right\rangle /2t\approx\varepsilon^{2}/2Vr_{m}^{2}.$
The transport efficiency reads as

\begin{equation}
\eta_{\theta}=\frac{v_{\theta}^{2}}{D_{\theta}\dot{W}}\approx\frac{2\omega_{s}}{\varepsilon^{2}\left(L_{12}-L_{21}\right)}.
\end{equation}

In the main text, we use the Eq.(\ref{eq:Stot}) allows us to numerically
calculate the exact dissipation rate $\dot{W}$ in Fig.\ref{fig:2}.
Since $\dot{x}_{i}$ and $H_{i}$ can be obtained from the dynamics
generating from Eq.(\ref{eq:CLE}), $\dot{s}_{m}$ can be calculated
numerically. By averaging over trajectories in steady states, $\dot{S}_{tot}$
and $\dot{W}$ can then be obtained.

\section{Pearson correlation coefficient}

In this section, we calculate the Pearson correlation coefficient
$\chi^{2}(R,Q)$ between the phase $R\left(\tau\right)=\theta\left(\tau\right)=\int_{0}^{\tau}\dot{\theta}\left(t\right)dt$
and the amplitude $Q\left(\tau\right)=\int_{0}^{\tau}r^{2}\left(t\right)dt$.
The change rate of the covariance, $C\left(r^{2},\theta\right)=\lim_{t\to\infty}\frac{1}{\tau}\text{Cov}_{r^{2},\theta}\left(\tau\right)$,
of these two variables can be calculated as

\begin{align}
\lim_{\tau\to\infty}\frac{1}{\tau}\text{Cov}(r^{n},\theta;\tau) & =\frac{1}{2\pi}\int_{0}^{2\pi}d\theta\int_{0}^{\infty}drr^{n}\dot{\theta}P_{ss}(r)\nonumber \\
 & \hphantom{}\hphantom{}-(\omega+C_{i}\left\langle r^{2}\right\rangle _{ss})\left\langle r^{n}\right\rangle _{ss}\nonumber \\
 & \approx C_{i}\left(\left\langle r^{n+2}\right\rangle _{ss}-\left\langle r^{n}\right\rangle _{ss}\left\langle r^{2}\right\rangle _{ss}\right).\label{eq:Cov-2}
\end{align}
Note that the integrals (averages) we are going to calculate all take
the form $I_{n}=\int_{0}^{\infty}r^{2n}\exp[V\varepsilon^{-2}(\frac{\alpha}{2}r^{2}+\frac{C_{r}}{4}r^{4})]dr^{2}=\int_{0}^{\infty}x^{n}\exp[V\varepsilon^{-2}(\frac{\alpha}{2}x+\frac{C_{r}}{4}x^{2})]dx$.
By setting $y=\rho(x/A-1)=\sqrt{-\frac{VC_{r}}{4\varepsilon^{2}}(x+\frac{\alpha}{C_{r}})}$
with $\rho=\alpha/2\sqrt{-C_{r}\varepsilon^{2}/V}$ and $A=-\alpha/C_{r}\approx r_{s}^{2}$
(for $\alpha>0$), we have
\begin{equation}
I_{n}=\exp\left(\rho^{2}\right)\left(r_{m}^{2}/\rho\right)^{n+1}\int_{-\rho}^{\infty}\left(y+\rho\right)^{n}e^{-y^{2}}dy.
\end{equation}
For $\alpha>0$, integrals $\int_{-\rho}^{\infty}p_{n}\left(y\right)e^{-y^{2}}$($p_{n}\left(y\right)$
are polynomials of degree $n$) can be obtained by simple Gaussian
integrals $\int_{-\infty}^{\infty}p_{n}\left(y\right)e^{-y^{2}}$,
for $\alpha=0$ integrals read $\int_{0}^{\infty}p_{n}\left(y\right)e^{-y^{2}}$
and for $\alpha<0$ integrals are approximately zero. For $\alpha>0$,
the covariance reads
\begin{equation}
C\left(r^{2},\theta\right)\approx C_{i}\left[r_{m}^{4}\left(1+\frac{1}{2\rho^{2}}\right)-\left(r_{m}^{2}\right)^{2}\right]=-\frac{2C_{i}C_{r}r_{m}^{4}\varepsilon^{2}}{\alpha^{2}V}.
\end{equation}
We need to emphasize that such equation holds in the region where
$\sqrt{-2C_{r}\varepsilon^{2}/V}<\alpha\ll\left|C_{r}/C_{i}\right|$
due the above approximation. Thus, for normal oscillation region ($\alpha>0$)
, the Pearson correlation coefficient $\chi^{2}\left(r^{2},\theta\right)$
can be calculated as
\begin{equation}
\chi^{2}\left(r^{2},\theta\right)\approx2\alpha(C_{i}/C_{r})^{2}<1,
\end{equation}
which means that such scheme works well for oscillators with high
value of $\left|C_{i}/C_{r}\right|$ (independent of the system size).

Generally, for $Q_{r,n}=\int_{0}^{\tau}r^{n}\left(t\right)dt$ with
$n$ the power of oscillatory amplitude, the change rate of the covariance
between oscillatory phase and amplitude is related to the higher-order
moment of amplitude as

\begin{align}
\lim_{t\to\infty}\frac{1}{\tau}\text{Cov}(r^{n},\theta;\tau) & =\left\langle \left\langle r^{n}\dot{\theta}\right\rangle \right\rangle _{ss}-\left\langle \left\langle \dot{\theta}\right\rangle \right\rangle _{ss}\left\langle \left\langle r^{n}\right\rangle \right\rangle _{ss}\nonumber \\
 & =\frac{1}{2\pi}\int_{0}^{2\pi}d\theta\int_{0}^{\infty}drr^{n}\dot{\theta}P_{ss}(r)\nonumber \\
 & ~-(\omega+C_{i}\left\langle r^{2}\right\rangle _{ss})\left\langle r^{n}\right\rangle _{ss}\nonumber \\
 & \approx C_{i}(\left\langle r^{n+2}\right\rangle _{ss}-\left\langle r^{n}\right\rangle _{ss}\left\langle r^{2}\right\rangle _{ss}).
\end{align}
and the Pearson correlation coefficient $\chi_{n}^{2}=\chi^{2}(r^{n},\theta)$
can be calculated as

\begin{align}
\chi_{n}^{2} & =\frac{\left[\text{Cov}(r^{n},\theta)\right]^{2}}{\text{Var}(r^{n})\text{Var}(\theta)}\nonumber \\
 & \approx\frac{C_{i}^{2}(\left\langle r^{n+2}\right\rangle _{ss}-\left\langle r^{n}\right\rangle _{ss}\left\langle r^{2}\right\rangle _{ss})^{2}}{D_{\theta}(\left\langle r^{2n}\right\rangle _{ss}-\left\langle r^{n}\right\rangle _{ss}^{2})}.
\end{align}

\section{Details of the models}

\subsection{The circadian clock model}

\begin{figure}
\begin{centering}
\includegraphics[width=1\columnwidth]{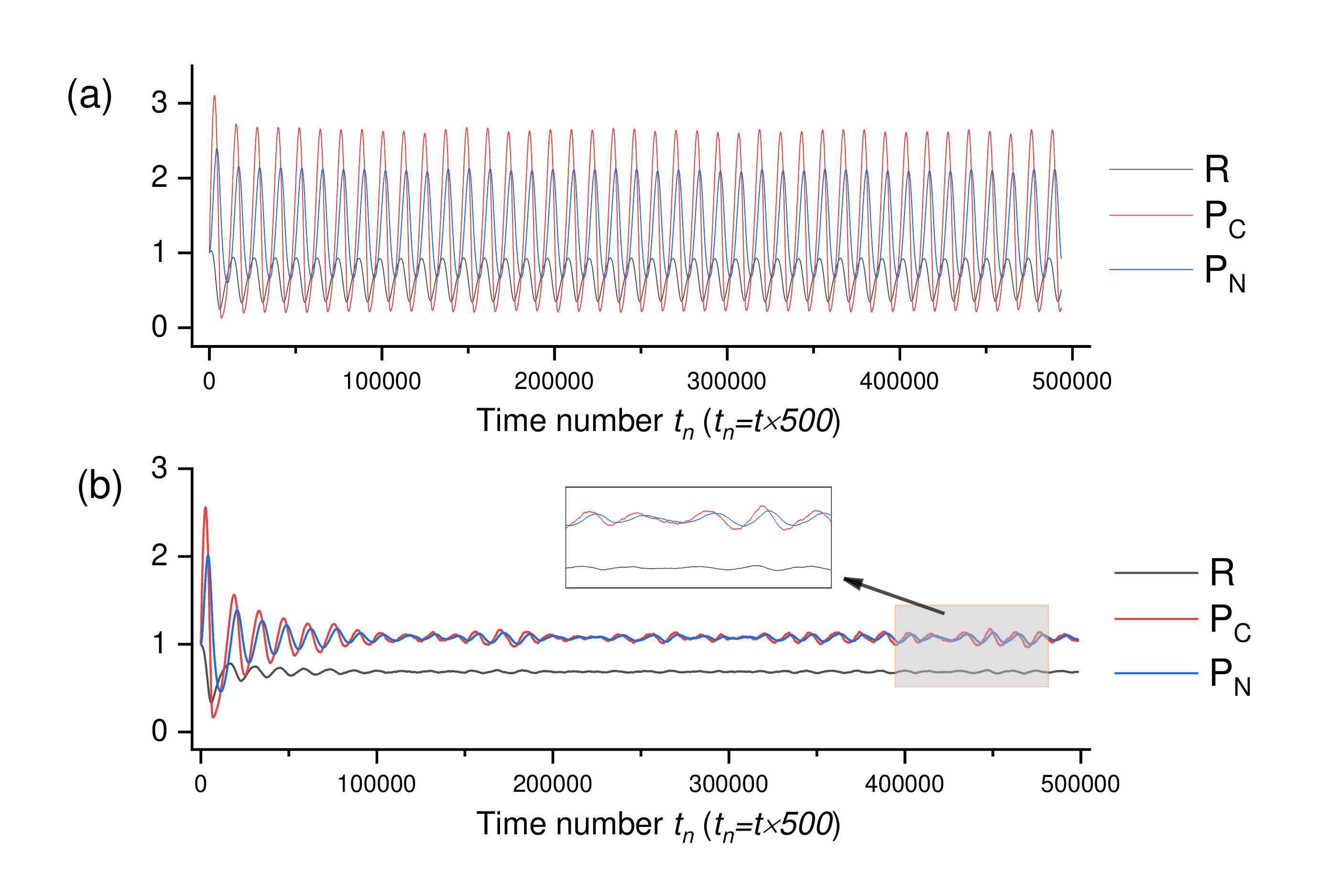}
\par\end{centering}
\caption{Typical trajectories for the circadian clock model. (a) $\alpha=0.05$.
(b) $\alpha=-0.01.$ The system size $V=1.6\times10^{5}$.}

\label{fig:5}
\end{figure}

Here, we describe the details of the circadian clock model studied
in the main text. The vector $\bm{x}=\left(x_{1},x_{2},x_{3}\right)$
stands for the concentrations of $\left(\text{R},\text{P}_{C},\text{P}_{N}\right)$.
The transcription rate of mRNA is chosen as the control parameter,
represented by $v_{s}$. As stated in the main text, the Hopf bifurcation
point locates at $v_{s}\simeq0.25725$. The the deterministic reaction
equations for the current model reads with $\bm{w}$ the transition
rates

\[
\text{\ensuremath{\frac{dx_{1}}{dt}=w_{1}-w_{2}}},
\]

\[
\text{\ensuremath{\frac{dx_{2}}{dt}=w_{3}-w_{4}-w_{5}+w_{6}}},
\]

\[
\text{\ensuremath{\frac{dx_{2}}{dt}=w_{5}-w_{6}}}.
\]

The descriptions of the reaction channels and values of parameters
are listed in Table I. Typical trajectories for the concentrations
of $\left(\text{R},\text{P}_{C},\text{P}_{N}\right)$ in this model
have been shown in Fig.\ref{fig:5}. In Fig.\ref{fig:6}(a), we plot
the the transport efficiencies as a function of the control parameter
$\alpha$ for the circadian clock model. It can be observed that the
estimator proposed by us yields a significant improvement over the
conventional TUR.
\begin{center}
{\scriptsize{}{}{}}%
\begin{longtable}[c]{|c||c||c||c|}
\caption{Descriptions of the circadian clock model}
\tabularnewline
\hline
 & {\scriptsize{}{}{}Reaction}  & {\scriptsize{}{}{}Transition rate}  & {\scriptsize{}{}{}Biochemical function}\tabularnewline
\hline
\hline
{\scriptsize{}{}{}1}  & {\scriptsize{}{}{}G $\to$ R + G}  & {\scriptsize{}{}{}$w_{1}=\frac{v_{s}k_{I}^{n}}{k_{I}^{n}+x_{3}^{n}}$}  & {\scriptsize{}{}{}Transcription}\tabularnewline
\hline
{\scriptsize{}{}{}2}  & {\scriptsize{}{}{}R $\to$}  & {\scriptsize{}{}{}$w_{2}=\frac{v_{m}x_{z}^{n}}{k_{m}+x_{1}}$}  & {\scriptsize{}{}{}R degradation}\tabularnewline
\hline
{\scriptsize{}{}{}3}  & {\scriptsize{}{}{}R $\to$ R + $\text{P}_{C}$}  & {\scriptsize{}{}{}$w_{3}=k_{s}x_{1}$}  & {\scriptsize{}{}{}Translation}\tabularnewline
\hline
{\scriptsize{}{}{}4}  & {\scriptsize{}{}{}$\text{P}_{C}$ $\to$}  & {\scriptsize{}{}{}$w_{4}=\frac{v_{d}x_{2}}{k_{d}+x_{2}}$}  & {\scriptsize{}{}{}Degradation of $\text{P}_{C}$}\tabularnewline
\hline
{\scriptsize{}{}{}5}  & {\scriptsize{}{}{}$\text{P}_{C}$ $\to$ $\text{P}_{R}$}  & {\scriptsize{}{}{}$w_{5}=k_{1}x_{2}$}  & {\scriptsize{}{}{}Transport of $\text{P}_{C}$ into the nucleus}\tabularnewline
\hline
{\scriptsize{}{}{}6}  & {\scriptsize{}{}{}$\text{P}_{N}$ $\to$ $\text{P}_{C}$}  & {\scriptsize{}{}{}$w_{6}=k_{2}x_{3}$}  & {\scriptsize{}{}{}Transport of $\text{P}_{N}$ out of the nucleus}\tabularnewline
\hline
\multicolumn{4}{|c|}{{\scriptsize{}{}{}$k_{I}=2.0$ nM, $n=4$, $v_{m}=0.3$ nM $\text{h}^{-1}$,
$k_{m}=0.2$ nM,}}\tabularnewline
\multicolumn{4}{|c|}{{\scriptsize{}{}{}$k_{s}=2.0$ $\text{h}^{-1}$, $v_{d}=1.5$ nM
$\text{h}^{-1}$, $k_{d}=0.1$ nM, $k_{1}=k_{2}=0.2$ $\text{h}^{-1}$}}\tabularnewline
\hline
\end{longtable}
\par\end{center}

\subsection{The Brusselator model }

\begin{figure}
\begin{centering}
\includegraphics[width=1\columnwidth]{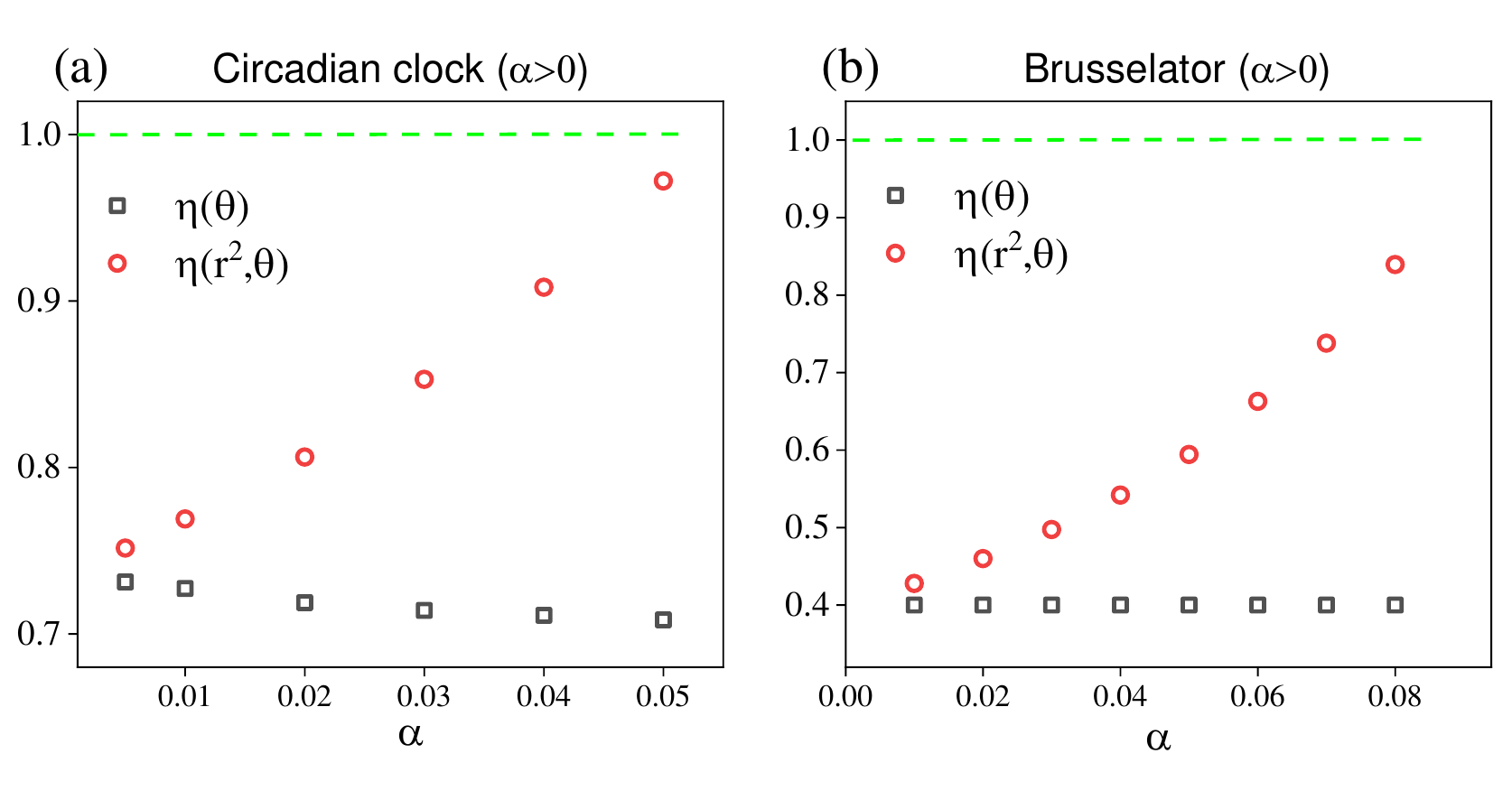}
\par\end{centering}
\caption{The transport efficiencies as a function of the control parameter
$\alpha$. The black squares correspond to the TUR, $\eta(\theta)=\dot{W}_{TUR}/\dot{W}$,
for the oscillatory phase $\theta$ only, and the red circles show
$\eta(r^{2},\theta)=\dot{W}_{I,2}/\dot{W}$ including the correlations
between the phase and amplitude $r$. (a) The circadian clock model.
(b) The Brusselator model. The system size $V=1.6\times10^{5}$.}

\label{fig:6}
\end{figure}

Here, we introduce the Brusselator model, involving two distinct biochemical
species $X$, $Y$, whose time evolution is governed by the following
deterministic kinetic equations:

\[
\frac{dX}{dt}=A-(B+1)X+X^{2}Y,
\]

\[
\frac{dY}{dt}=BX-X^{2}Y.
\]
In the deterministic limit, the system has a stable point $X_{s}=A$,
$Y_{s}=B/A$, which loses stability when the control parameter $B$
exceeds the Hopf bifurcation point $B_{c}=1+A^{2}$. The normal biochemical
oscillation happens for $B>B_{c}$. By choosing $A=0.3$, we calculate
the parameters in stochastic normal form theory as $C_{r}\simeq-2.9028$,
$C_{i}\simeq5.2506$, $\varepsilon^{2}=4$, $\omega=1$, and $\alpha=(B-1-A^{2})/2$.
In Fig.\ref{fig:6}(b), the the transport efficiencies as a function
of the control parameter $\alpha$ for the Brusselator model have
been depicted. As expected, the improved estimator proposed by us
is much more accurate than the TUR bound.

 \bibliographystyle{apsrev}

\end{document}